# A Multidimensional Evaluation Framework for Artificial Intelligence-Based Ocean State Forecasts

Ziqing ZU[1,2], Ye YUAN *[1,2], Tiejun LING[1,2], Zhi CHEN[1,2], Xiang LI[1,2], Zhiqiang LI[1,2], Qiang WANG[3,4], Jiangjiang XIA[5,6], Jiuke WANG[7], Yongjun JIA[8], Shinan WANG[1,2], Zhi-yuan ZHANG[9], Renbo PANG[1,2], Xiaojuan LIN[1,2], Yu ZHANG[1,2], Huier MO[1,2], Ang LI[1,2]

[1] *State Key Laboratory of Satellite Ocean Environment Dynamics, National Marine Environmental Forecasting Center, Beijing 100081, China;*

[2] *Key Laboratory of Marine Hazards Forecasting, Ministry of Natural Resources, Beijing 100081, China;*

[3] *Key Laboratory of Marine Hazards Forecasting, Ministry of Natural Resources, Hohai University, Nanjing 210098, China;*

[4] *College of Oceanography, Hohai University, Nanjing 210098, China;*

[5] *Temperate East Asia Research Center on Global Change, Institute of Atmospheric Physics, Chinese Academy of Sciences, Beijing 100029, China;*

[6] *University of Chinese Academy of Sciences, Beijing 100049, China;*

[7] *School of Artificial Intelligence, Sun Yat-sen University, Zhuhai 519082, China;*

[8] *National Satellite Ocean Application Service, Beijing 100081, China;*

[9] *Unit No.91001 of PLA, Beijing 100036, China.*

* Corresponding author: yuanye@nmefc.cn

## Abstract

Artificial intelligence (AI)-based ocean state forecasts now approach or surpass those produced by operational ocean forecasting systems as measured by conventional pointwise metrics. However, these metrics may favor spatially smoothed forecasts and cannot determine whether forecast variables remain physically coherent. Forecast evaluation should therefore move beyond overall pointwise accuracy toward assessing forecast trustworthiness. We propose a multidimensional framework comprising overall pointwise accuracy, physical consistency, and distributional consistency. Overall pointwise accuracy summarizes forecast–reference agreement over all valid samples; physical consistency evaluates dynamical balances using model outputs alone; and distributional consistency evaluates distribution tails and multiscale spatial structures. Three AI models were deployed as representative testbeds in a unified operational environment. This comprehensive assessment shows that lower overall pointwise errors cannot preclude substantial dynamical imbalances and suppressed fine-scale variance. Conversely, higher pointwise errors may arise from spatial displacement of coherent structures rather than a poor representation of their amplitude. The results demonstrate that the evaluation of AI ocean forecasting models should be based on the multi-dimensional assessment of forecast trustworthiness, rather than solely relying on a single metric. The study provides a preliminary framework for AI model intercomparison, which is also informative for targeted model refinement and operational implementation.

**Keywords**: AI ocean forecasting; forecast evaluation; physical consistency; distributional consistency; multiscale spatial structures; extreme events

**Plain Language Summary**

AI models are increasingly used to forecast ocean state variables, and they are often judged using the average differences between forecasts and observations. A small average error, however, does not necessarily mean that a forecast is trustworthy. A model may achieve a low overall forecast error by smoothing eddies and fronts, while its forecast variables may not remain physically consistent. We therefore developed a framework that examines forecasts in three dimensions: their overall agreement with observations, their internal physical coherence, and their ability to forecast extreme values and retain the fine-scale structures. We applied this framework to three AI models deployed in the same operational environment. Forecast errors were generally larger for extreme values than when all values were considered. A low average error could coexist with overly smoothed fine-scale structures and physically inconsistent relationships among forecast variables, whereas a larger error could result from a coherent ocean feature being slightly misplaced. These results show why no single metric can measure forecast trustworthiness. All three dimensions: overall pointwise accuracy, physical consistency, and distributional consistency, should be evaluated together.

## 1 Introduction

Synoptic-scale ocean state forecasting estimates the evolution of temperature, salinity, currents, and sea surface height over lead times of up to two weeks, supporting marine hazard preparedness, resource management, and environmental services. In the past five years, the accumulation of ocean data, together with advances in computational resources and deep-learning algorithms, has enabled the development of artificial intelligence (AI)-based ocean state forecasting models (Wang et al., 2024; Cui et al., 2025; El Aouni et al., 2025a; Yang et al., 2026). These models have shown performance comparable to or better than that of operational ocean forecasting systems when evaluated using conventional pointwise metrics such as root-mean-square error (RMSE) (Zu et al., 2025). A fundamental question is whether AI forecasts also reproduce the physical coherency and distributional characteristics of ocean states that are not fully captured by conventional pointwise metrics.

Favorable values of conventional pointwise metrics do not necessarily imply physical consistency. Most AI forecasting models learn statistical relationships without explicitly enforcing conservation laws or dynamical balances, and are trained on datasets that inevitably contain uncertainties. Conventional pointwise metrics such as RMSE quantify forecast–reference differences, but cannot determine whether forecast variables jointly satisfy these physical constraints. Recent studies have shown that Pangu-Weather fails to close the global atmospheric mass budget, whereas GraphCast exhibits anomalous variations in atmospheric water vapor (Bonavita, 2024). Such physical inconsistencies can undermine forecast trustworthiness and limit the use of AI forecasts in mechanistic interpretation, data assimilation, and predictability studies.

Metrics for overall pointwise accuracy also offer limited insight into forecast performance in the distribution tails and multiscale spatial structures. Loss functions such as mean squared error (MSE) and evaluation metrics such as RMSE are dominated by the non-extreme samples, so favorable overall metric values may coexist with substantially larger errors under extreme conditions. Additionally, minimizing MSE encourages forecasts toward the conditional mean, which can suppress fine-scale variance and produce overly smoothed spatial structures (Husain et al., 2025). Overall pointwise accuracy alone is therefore insufficient to assess forecast performance for extreme-event early warning or to assess the representation of mesoscale processes.

These concerns have motivated AI forecast evaluation to expand beyond overall pointwise accuracy by introducing metrics for physical consistency and multiscale spatial structures. In atmospheric forecasting, WeatherBench and WeatherBench2 established standardized evaluation frameworks using common datasets and metrics (Rasp et al., 2020, 2024). ChaosBench further extended the evaluation to subseasonal-to-seasonal prediction and introduced physics-based spectral metrics (Nathaniel et al., 2024). In ocean forecasting, OceanForecastBench similarly established a standardized dataset and evaluation workflow, although it primarily uses pointwise metrics (Han et al., 2026). OceanBench broadened the evaluation scope by introducing additional metrics, including mixed-layer depth, geostrophic currents, Lagrangian trajectories, and effective resolution (El Aouni et al., 2025b).

Despite these advances, three gaps remain in current evaluations of AI ocean forecasts. First, the existing benchmarks are designed for offline evaluation using fixed historical datasets. The coverage and sampling of these reference datasets may limit the assessment of end-to-end forecast performance and generalization under operational conditions. Second, the physical metrics introduced in OceanBench quantify errors in derived variables with respect to the reference data, rather than the internal physical consistency among forecast variables. Third, existing frameworks have not systematically assessed how well AI forecasts represent distribution tails and multiscale spatial structures. This limits their ability to distinguish errors associated with extreme events, fine-scale structures, and structural displacements.

To address these gaps, we propose a multidimensional evaluation framework comprising overall pointwise accuracy, physical consistency, and distributional consistency. The objective is not to identify a universally best model, but to evaluate complementary aspects of forecast trustworthiness. Wenhai, Skyocean, and ST-base are deployed as representative testbeds in a unified operational environment. Variations in their neural network architecture, training data, input data, and operational configurations allow us to examine whether the proposed metrics can reveal forecast deficiencies overlooked by overall pointwise metrics, particularly with regard to physical consistency and the representation of extreme values and fine-scale structures.

## 2 Models, data and evaluation configurations

### 2.1 Models

Forecasts generated by three AI models are evaluated in this study. The models were deployed in a unified operational forecasting environment to produce continuous rolling forecasts for the same initialization dates and lead times. Each model retained its original input data and operational configuration, which are summarized in Table 1.

Wenhai, jointly developed by Laoshan Laboratory and its partner institutions, is a global eddy-resolving AI ocean forecasting model with a horizontal resolution of 1/12° (Cui et al., 2025). It was trained on GLORYS and ERA5 data from 1993 to 2018. For operational forecasting, Wenhai's atmospheric forcing is calculated using bulk formulas (Large and Yeager, 2009) and variables from atmospheric forecasts throughout the forecast period. It produces daily mean forecasts of temperature, salinity, currents, and SSH for lead times of up to 15 days.

Skyocean, jointly developed by the National Satellite Ocean Application Service and Sun Yat-sen University, is a regional AI ocean forecasting model that integrates satellite remote sensing, ocean state reconstruction, and short-term forecasting. It was trained using temperature, salinity, and currents from GLORYS, and SLA from AVISO L4. For operational forecasts, Skyocean uses an ocean analysis from the day before initialization, supplemented by SSS from SMAP and SMOS, SST from OISST, and SLA from AVISO L4. The model covers the northwestern Pacific at a horizontal resolution of 1/8° and produces forecasts of temperature, salinity, currents, and SLA for lead times of up to 10 days.

ST-base, developed by the National Marine Environmental Forecasting Center, is a regional AI ocean forecasting model designed for operational testing in the northwestern Pacific. It employs a hierarchical attention architecture based on the Swin Transformer and has a horizontal resolution of 1/12° with 50 vertical levels. For each initialization, ST-base uses daily ocean and atmospheric analysis for the seven-day period before initialization. It produces forecasts of temperature, salinity, currents, and SSH for lead times of up to 7 days. The model uses a baseline training configuration without specialized training strategies and serves as the baseline model in this study.

Table 1. Input data and operational configurations of the three AI-based ocean forecasting models.

| Model | Ocean inputs | Atmospheric inputs | Notes |
|---|---|---|---|
| Wenhai | Analysis from the day before initialization: temperature, salinity, horizontal velocities, SSH. | Forecasts throughout the forecast period: 10-m winds, 2-m air temperature, 2-m dew-point temperature, surface downward longwave radiation, net surface shortwave radiation, precipitation, and mean sea level pressure. | Wenhai was fine-tuned for an operational application; air–sea fluxes were calculated using bulk formulas. |
| Skyocean | Analysis from the day before initialization: temperature, salinity, horizontal velocities, supplemented with SSS from SMAP and SMOS, SST from OISST, and SLA from AVISO L4. | No atmospheric inputs. | AVISO L4 SLA replaced SSH analysis. |
| ST-base | Analysis for the seven-day period before initialization: temperature, salinity, horizontal velocities and SSH. | Analysis for the seven-day period before initialization: 2-m air temperature and 10-m winds. | Baseline configuration without specialized training strategies |

Note. AVISO: Archiving, Validation, and Interpretation of Satellite Oceanographic data; OISST: Optimum Interpolation Sea Surface Temperature; SLA: sea level anomaly; SMAP: Soil Moisture Active Passive; SMOS: Soil Moisture and Ocean Salinity; SSH: sea surface height; SSS: sea surface salinity; SST: sea surface temperature; L4: Level 4.

Each AI model is combined with its model-specific input data and operational configuration, and is thus treated as an integrated forecasting system. The forecast errors therefore reflect the combined effects of the model and its inputs, rather than network architecture alone. Wenhai uses atmospheric forecasts as forcing over the forecast period, which can improve its forecast skill. This is common practice in operational ocean forecasting and is therefore appropriate for the system-level evaluation adopted in this study.

## 2.2 Data

Overall pointwise accuracy is evaluated against multiple observational datasets. Temperature and salinity forecasts are evaluated against Argo profiles, SLA forecasts against AVISO L3 along-track altimeter observations, and horizontal current forecasts against 15-m velocities derived from surface drifters. These reference datasets are provided in a quality-controlled and standardized format through the OceanPredict Intercomparison and Validation Task Team (IV-TT) Class 4 project (Ryan et al., 2015).

The evaluation of distributional consistency is limited to SST and SLA because only their corresponding reference datasets provide adequate spatial coverage and resolution for analysis of distribution tails and multiscale structures. SST forecasts are evaluated against SST references from the Operational Sea Surface Temperature and Ice Analysis (OSTIA), whereas SLA forecasts are evaluated against SLA references from AVISO L4 and Surface Water and Ocean Topography (SWOT) L3 wide-swath altimeter observations.

Physical consistency is assessed by examining dynamical balances among the forecast variables rather than by comparison with external reference datasets. A climatology of the Rossby deformation radius associated with the first surface mode, calculated from World Ocean Atlas 2018, is used solely to define the spatial filtering scale for the geostrophic-balance metric (LaCasce and Groeskamp, 2020).

Mean dynamic topography (MDT) is used to convert between SSH and SLA. For SLA-based comparisons, GLORYS MDT is subtracted from the SSH forecasts produced by Wenhai and ST-base. Conversely, for the geostrophic-balance metric, Global Ocean MDT from the Copernicus Marine Environment Monitoring Service (CMEMS) is added to Skyocean's SLA forecasts to obtain SSH.

### 2.3 Evaluation configurations

The evaluation domain covers the northwestern Pacific (10°–35°N, 121°–150°E). Characterized by strong currents, sharp oceanic fronts, and energetic mesoscale eddies, this region provides a suitable testbed for evaluating multiscale spatial structures and dynamical consistency (Chelton et al., 2011). The domain is located within the overlapping interior of the three model domains, thereby minimizing the influence of model boundaries. The evaluation period extends from 1 January to 30 April 2026 and comprises 120 daily forecast initializations. Forecast lead times range from 1 to 7 days, and evaluations of subsurface variables are restricted to the common depth range of the three models (0–644 m).

## 3 Evaluation framework

The framework evaluates forecasts across three complementary dimensions: overall pointwise accuracy, physical consistency, and distributional consistency. Overall pointwise accuracy summarizes forecast–reference agreement across all valid samples using conventional pointwise metrics. Physical consistency examines whether forecast variables satisfy dynamical balances, whereas distributional consistency assesses forecast errors for the distribution tails and multiscale spatial structures.

### 3.1 Overall pointwise accuracy

Overall pointwise accuracy is evaluated following the OceanPredict IV-TT Class 4 framework, in which model forecasts are assessed in observation space (Ryan et al., 2015). It uses conventional overall metrics calculated from the original, unfiltered forecast and reference fields across all valid samples. For each initialization date and lead time, the forecasts are interpolated to the horizontal locations, depths, and valid times of the observations. Only forecast–reference pairs satisfying the prescribed spatial and temporal collocation criteria are retained to calculate pointwise metrics.

For each lead time, bias, RMSE, and the anomaly correlation coefficient (ACC) are calculated from the retained forecast–reference pairs. Bias measures systematic errors. RMSE summarizes the magnitude of pointwise forecast errors. ACC measures the correspondence between forecast and reference anomalies relative to climatology. These metrics are defined as:

$$Bias_l = \frac{1}{N}\sum_{p=1}^{N}\left(f_{p,l} - r_p\right) \tag{1}$$

$$RMSE_l = \sqrt{\frac{1}{N}\sum_{p=1}^{N}\left(f_{p,l} - r_p\right)^2} \tag{2}$$

$$ACC_l = \frac{\sum_{p=1}^{N}\left(f_{p,l} - c_p\right)\left(r_p - c_p\right)}{\sqrt{\sum_{p=1}^{N}\left(f_{p,l} - c_p\right)^2 \sum_{p=1}^{N}\left(r_p - c_p\right)^2}} \tag{3}$$

where $f_p$ , $r_p$ and $c_p$ denote the interpolated forecast, reference, and climatology, respectively, for the $p$th paired sample, $l$ is lead day and $N$ is the number of retained pairs.

### 3.2 Physical consistency

Physical consistency is assessed directly from model outputs rather than through comparison with external references. Two metrics are used to evaluate complementary aspects of internal physical coherence: geostrophic balance and stratification stability.

#### 3.2.1 Stratification stability

The stratification stability metric evaluates whether the forecast density profiles remain stable. Stratification stability is quantified using the squared buoyancy frequency,

$$N_{i,j,k}^2 = \frac{g}{\rho_0}\frac{\rho_\theta^{i,j,k+1} - \rho_\theta^{i,j,k}}{\Delta z_{i,j,k}} \tag{4}$$

where $g$ is gravitational acceleration, $\rho_0$ is the reference density, and $\rho_\theta^{i,j,k}$ is the potential density at location $(i,j,k)$. The vertical index $k$ increases downward, and $\Delta z_{i,j,k} > 0$ is the layer thickness. Therefore, $N_{i,j,k}^2 > 0$ indicates stable stratification.

Because of interpolation errors, daily-mean forecasts may exhibit weak stratification instabilities. Based on an examination of daily-mean outputs from a conventional numerical ocean forecasting system in the study region, we adopt $N^2 \leq -10^{-5}$ $s^{-2}$ as the criterion for strong instability. The fraction of the total water volume that is strongly unstable is then defined as the stratification-stability metric:

$$F = \frac{\sum_{i,j,k} \delta_{i,j,k} \Delta V_{i,j,k}}{\sum_{i,j,k} \Delta V_{i,j,k}} \tag{5}$$

$$\delta_{i,j,k} = \begin{cases} 1, & N^2_{i,j,k} \leq -10^{-5}\ s^{-2} \\ 0, & otherwise \end{cases} \tag{6}$$

Here, $\Delta V_{i,j,k}$ is the volume of grid cell $(i,j,k)$. A smaller $F$ indicates more stable stratification in the forecasts.

### 3.2.2 Geostrophic balance

The geostrophic-balance metric examines the dynamical consistency between SSH and surface currents at geostrophic scales. The geostrophic scale varies with latitude (Vallis, 2017). A latitude-dependent low-pass filter is therefore applied to the SSH and surface currents. The cutoff wavelength of the filter is defined as $\lambda_c = 2L_R$, where $L_R$ is the first surface-mode deformation radius derived from World Ocean Atlas 2018 (LaCasce and Groeskamp, 2020).

To reduce the influence of regions where relative vorticity is large compared with the Coriolis parameter, particularly at low latitudes, a weight is assigned to each grid point:

$$W = |f|/(|f| + |\bar{\zeta}| + \varepsilon) \tag{7}$$

where $\bar{\zeta}$ is the relative vorticity calculated from the low-pass-filtered surface currents, $\varepsilon$ is a small positive constant set to $10^{-6}$ $s^{-1}$. The weighted root-mean-square geostrophic imbalance is defined as:

$$RMSE_{geo} = \left\{ \frac{\sum_{i,j} W_{i,j} \left[ \left( \overline{u}_{i,j} - \overline{u}_{g,i,j} \right)^2 + \left( \overline{v}_{i,j} - \overline{v}_{g,i,j} \right)^2 \right]}{\sum_{i,j} W_{i,j}} \right\}^{1/2} \tag{8}$$

where $\overline{u}_{i,j}$ and $\overline{v}_{i,j}$ are the low-pass-filtered large-scale surface currents, and $\overline{u}_{g,i,j}$ and $\overline{v}_{g,i,j}$ are the geostrophic currents derived from the low-pass-filtered SSH. A smaller $\mathrm{RMSE}_{geo}$ indicates better geostrophic balance.

### 3.3 Distributional consistency

Distributional consistency evaluates forecast performance in the tails of the reference-value distribution and across spatial scales. Three metrics are considered: extreme-event forecast errors, power spectral density (PSD), and scale-decomposed forecast errors. Because of limitations in reference coverage and spatial resolution, distributional consistency is evaluated only for SST and SLA using the reference datasets described in Section 2.2.

#### 3.3.1 Extreme-event forecast errors

Extreme-event forecast errors quantify forecast performance in the lower and upper tails of the reference distribution. Because ocean state variables exhibit pronounced seasonal cycles and long-term trends, dynamic percentile thresholds are constructed using 30 years of historical daily data. Let $X_{t,i,j}$ denote the historical reference at time $t$ and grid point $(i, j)$, and let $d_t$ denote the corresponding day of year from 1 to 365. At each grid point $(i, j)$, a linear trend is fitted for each day of year $d_t$ and then removed,

$$X'_{t,i,j} = X_{t,i,j} - a_{d_t,i,j} - b_{d_t,i,j}(t - \tau_0) \tag{9}$$

where $\tau_0$ is the start time of the historical dataset, $a_{d_t,i,j}$ and $b_{d_t,i,j}$ are the intercept and trend that vary with day of year and location.

To obtain stable estimates of the percentile thresholds, residuals $X'_{t,i,j}$ within an 11-day window are pooled to calculate the 10th and 90th percentiles,

$$Q_{p,d_t,i,j} = \mathrm{Quantile}_p\{X'_{t,i,j} \mid d_t \in \mathcal{W}_d\},\ p \in \{10, 90\} \tag{10}$$

where $\mathcal{W}_d$ denotes the 11-day window centered on day $d_t$. The dynamic percentile threshold at time $t$ is then,

$$T_{p,t,i,j} = a_{d_t,i,j} + b_{d_t,i,j}(t - \tau_0) + Q_{p,d_t,i,j} \tag{11}$$

For 29 February in leap years, the dynamic percentile threshold is obtained by linear interpolation between the thresholds for 28 February and 1 March.

Lower- and upper-tail extreme samples are identified using

$$\begin{aligned} \delta^{-}_{t,i,j} &= \begin{cases} 1, & r_{t,i,j} < T_{10,t,i,j}, \\ 0, & \text{otherwise}, \end{cases} \\ \delta^{+}_{t,i,j} &= \begin{cases} 1, & r_{t,i,j} > T_{90,t,i,j}, \\ 0, & \text{otherwise}. \end{cases} \end{aligned} \tag{12}$$

where $r_{t,i,j}$ denotes the reference at time $t$ and location $(i, j)$. The RMSE for each class of extreme samples is calculated as

$$RMSE_{e,l} = \sqrt{\frac{\sum_{t,i,j} \delta^{e}_{t,i,j} \left(f_{l,t,i,j} - r_{t,i,j}\right)^2}{\sum_{t,i,j} \delta^{e}_{t,i,j}}}, \quad e \in \{-, +\} \tag{13}$$

For SST, the dynamic thresholds are constructed from historical OSTIA data (1992-2021). For SLA, the thresholds are constructed from historical AVISO L4 data (1993-2022), while forecast errors are calculated against AVISO L4 and SWOT L3, respectively.

### 3.3.2 Power spectral density

PSD quantifies how variance is distributed across spatial scales and can reveal scale-dependent deficits in forecast variance (Skamarock, 2004). For SST, spatial means are removed from the forecasts and OSTIA references, followed by two-dimensional application of a Hann window. PSDs are then calculated using a two-dimensional fast Fourier transform. For SLA, forecasts are first interpolated to the AVISO L4 and SWOT L3 locations. One-dimensional meridional PSDs are then calculated for both forecasts and AVISO L4, while along-track PSDs are calculated for both forecasts and SWOT L3.

### 3.3.3 Scale-decomposed forecast errors

Scale-decomposed forecast errors are used to quantify forecast–reference differences across spatial scales. To avoid confusion with commonly used terminology, scales A, B, and C are defined as wavelength bands greater than 300 km, between 70 and 300 km, and less than 70 km, respectively. Gaussian spatial filters are then applied to extract the corresponding components of the reference and forecast. The standard deviation of the Gaussian kernel is,

$$\sigma_\lambda = \lambda/(2\pi) \tag{14}$$

where the cutoff wavelength $\lambda$ is set to 300 or 70 km. Scale-decomposed RMSE measures pointwise agreement at individual spatial scales:

$$\mathrm{RMSE}_{s,l} = \sqrt{\frac{1}{N_{s,l}}\sum_{p=1}^{N_{s,l}}\left(f_{s,l,p} - r_{s,l,p}\right)^2}, \quad s \in \{A, B, C\} \tag{15}$$

where $N_{s,l}$ is the grid number of valid forecast–reference pairs at the scale $s$ and lead time $l$.

RMSE is sensitive to both structural amplitude and displacement errors. A displaced eddy or front can produce a large RMSE even when its spatial structure is well represented. To reduce this double penalty, we additionally use the neighborhood continuous ranked probability score (CRPS; Ebert, 2008; Stein and Stoop, 2022). Following Cui et al. (2025), the neighborhood CRPS is defined as:

$$CRPS_{s,l} = \frac{1}{N_{s,l}}\sum_{p=1}^{N_{s,l}}\int_{-\infty}^{\infty}[F(f_{s,l,p}) - H(f_{s,l,p} \geq r_{s,p})]^2 d\, f_{s,l,p} \tag{16}$$

where $H$ is the Heaviside step function, equal to 1 when the integration variable $f_{s,l,p}$ is greater than or equal to the reference and 0 otherwise. The forecast values within a neighborhood ($1° \times 1°$) of each reference location $p$ are used to construct a local cumulative distribution function $F(f_{s,l,p})$. A smaller CRPS indicates closer agreement between the local distribution of forecast values and the reference value.

Scale-decomposed RMSE measures pointwise agreement at a specific scale, whereas neighborhood CRPS is less sensitive to small spatial displacement (Gilleland et al., 2009). Together with PSD, these metrics help determine whether scale-dependent errors are associated with spatial displacement or deficient fine-scale variance.

## 4 Results

### 4.1 Overall pointwise accuracy

Figure 1 shows that overall pointwise accuracy depends on both forecast variable and lead time. For temperature, the Wenhai and Skyocean curves intersect after lead day 3 for both RMSE and ACC, reflecting differences in their lead-time evolution. Salinity exhibits a similar transition near lead day 6. The temperature RMSE of ST-base rises more steeply with lead time. All three forecasts exhibit positive temperature and salinity biases, indicating systematic warm and saline departures from the reference fields. These results show why lead-time evolution, rather than a lead-averaged metric value, is needed to characterize forecast performance of temperature and salinity.

For SLA, differences in RMSE among the three forecasts persist throughout the forecast period. All three forecasts exhibit negative biases. The lower errors of the Skyocean forecasts against AVISO L3, in terms of both RMSE and absolute bias, may be partly attributed to its use of AVISO L4 SLA during training and inference. This potential reference dependence is examined further using independent references and scale-dependent metrics in Section 4.3.

For horizontal currents at 15 m, the RMSEs of both zonal (U) and meridional (V) velocities remain between approximately 0.21 and 0.235 m $s^{-1}$, with relative differences among the three forecasts generally below 5%. All three forecasts exhibit positive biases in both components, while ACC generally decreases with lead time. These results indicate that the three forecasts have broadly similar error characteristics, with only small differences that vary with lead time.

### 4.2 Physical consistency

Figure 2 presents two physical-consistency metrics: stratification stability and geostrophic balance. Unlike the overall pointwise accuracy metrics that are evaluated against references, these metrics are calculated directly from model outputs and assess the internal physical consistency of each forecast.

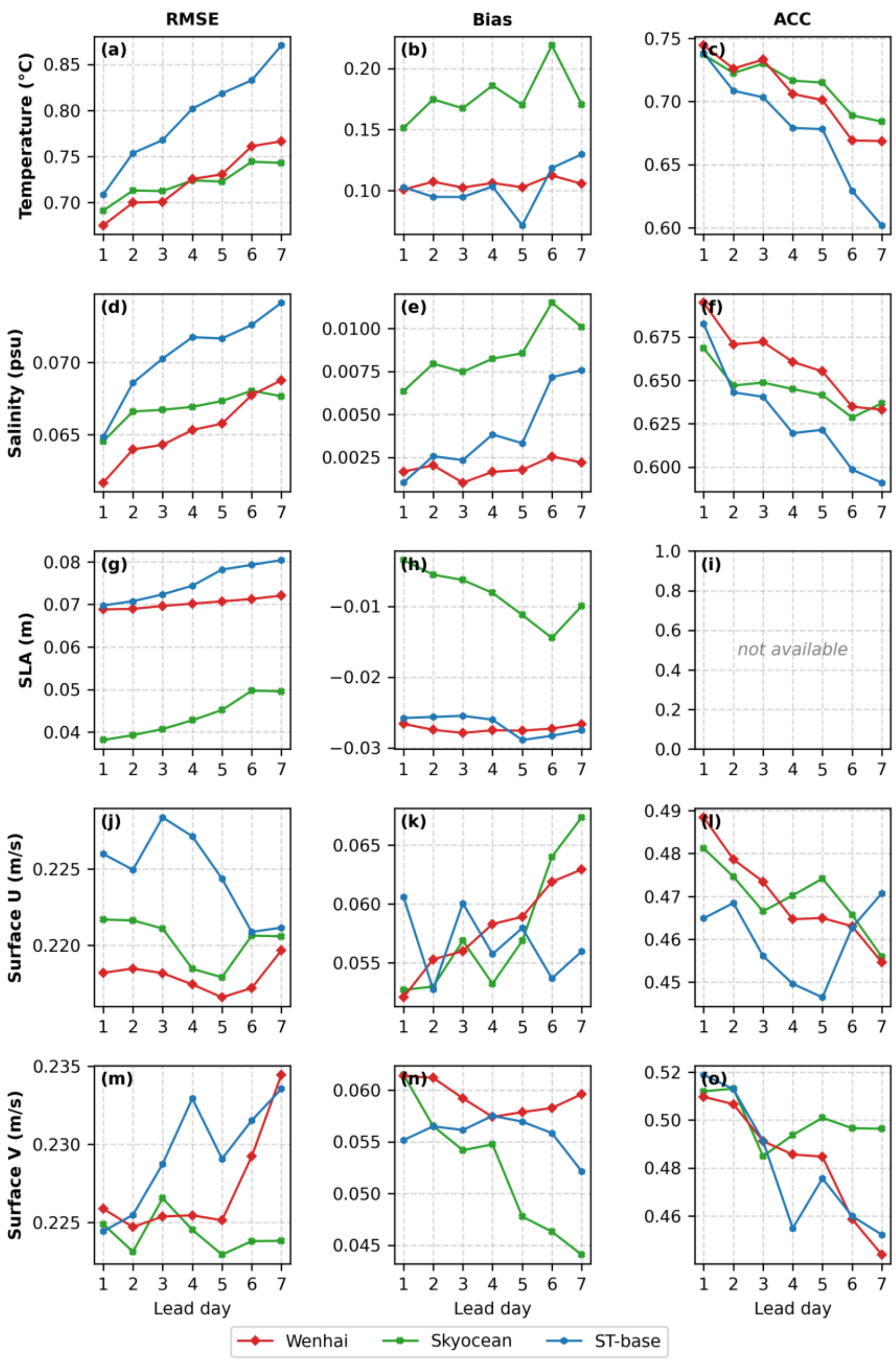

Figure 1. Overall pointwise accuracy of the three forecasts with lead time. ACC is not reported for SLA because no climatology is available.

Stratification stability evolves differently among the three forecasts (Figure 2a). The fraction of unstable water volume in Wenhai increases from approximately 0.5% on lead day 1 to 2.7% on lead day 7. The fraction in Skyocean remains between 1.0% and 1.5%, whereas that in ST-base decreases from approximately 3.3% to 2.6%. This comparison shows that pointwise metrics cannot indicate whether the fraction of strongly unstable water increases, remains stable, or decreases over the forecast lead time (Figures 1a-f).

The geostrophic-balance residual quantifies the consistency between SSH and surface currents (Figure 2b). The residual for Wenhai remains nearly constant at approximately 0.10 m $s^{-1}$, while that for ST-base ranges from approximately 0.107 to 0.118 m $s^{-1}$. For Skyocean, the residual decreases from approximately 0.176 m $s^{-1}$ on lead day 1 to 0.138 m $s^{-1}$ on lead day 7. Although Skyocean exhibits relatively small pointwise SLA errors (Figure 1g-h), its geostrophic-balance residual remains substantial. This contrast further demonstrates that low pointwise SLA errors alone do not ensure dynamical consistency between SSH and surface currents at geostrophic scales.

The two metrics show that different aspects of physical consistency do not necessarily vary together. Relatively stable stratification can coexist with a large geostrophic-balance residual, and vice versa.

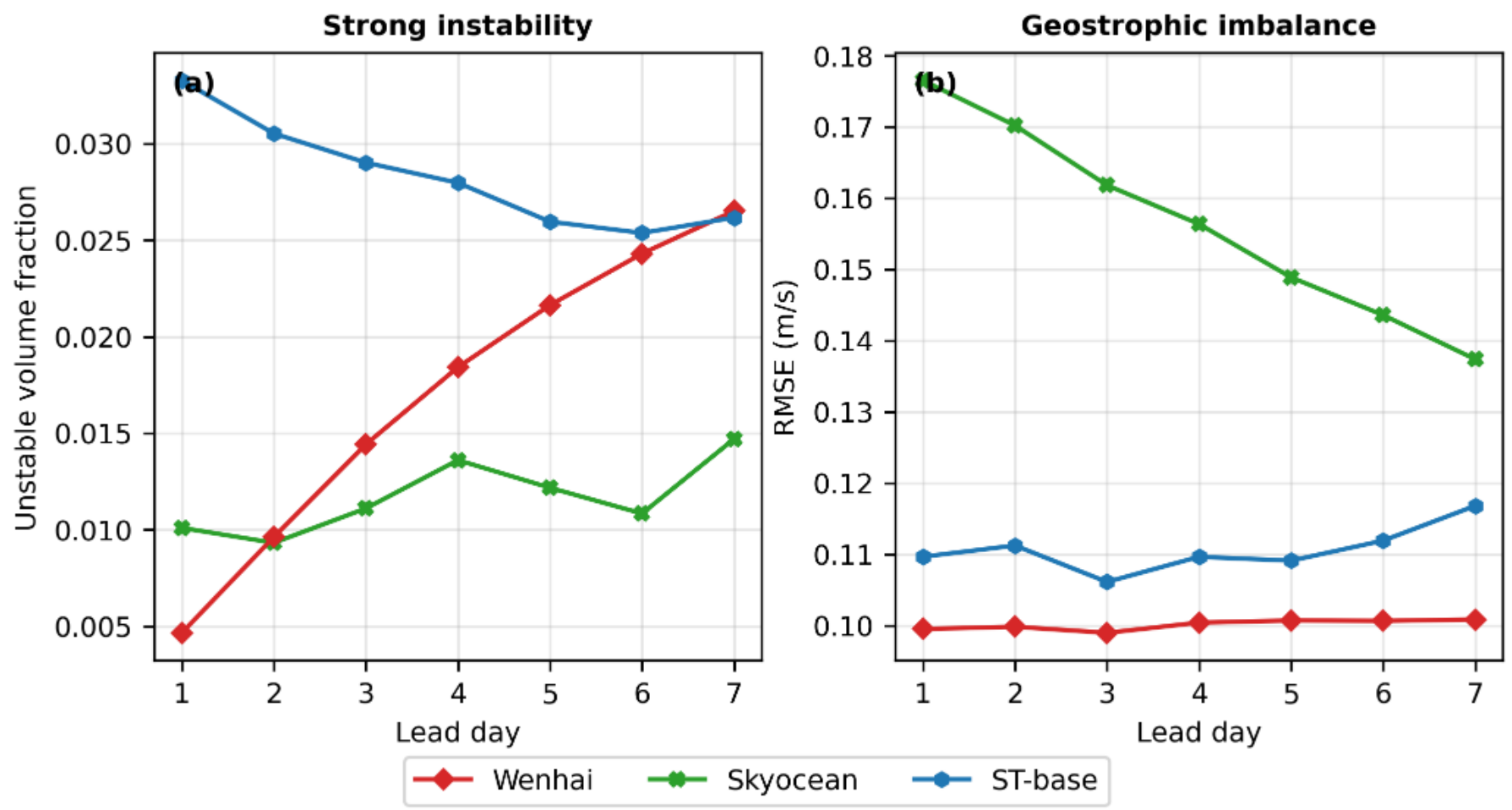

Figure 2. Evolution of physical-consistency metrics with lead time: (a) fraction of strongly unstable ($N^2 \leq -10^{-5}$ $s^{-2}$) water volume, (b) geostrophic-balance residual.

## 4.3 Distributional consistency

### 4.3.1 Extreme-event forecast errors

Figure 3 compares RMSEs for negative extremes, positive extremes, and the full sample for the three forecasts. Across the variables and references, extreme-event

RMSEs are generally higher than the full-sample RMSEs. Because non-extreme cases constitute most of the samples, full-sample RMSEs can mask the larger forecast errors in the distribution tails.

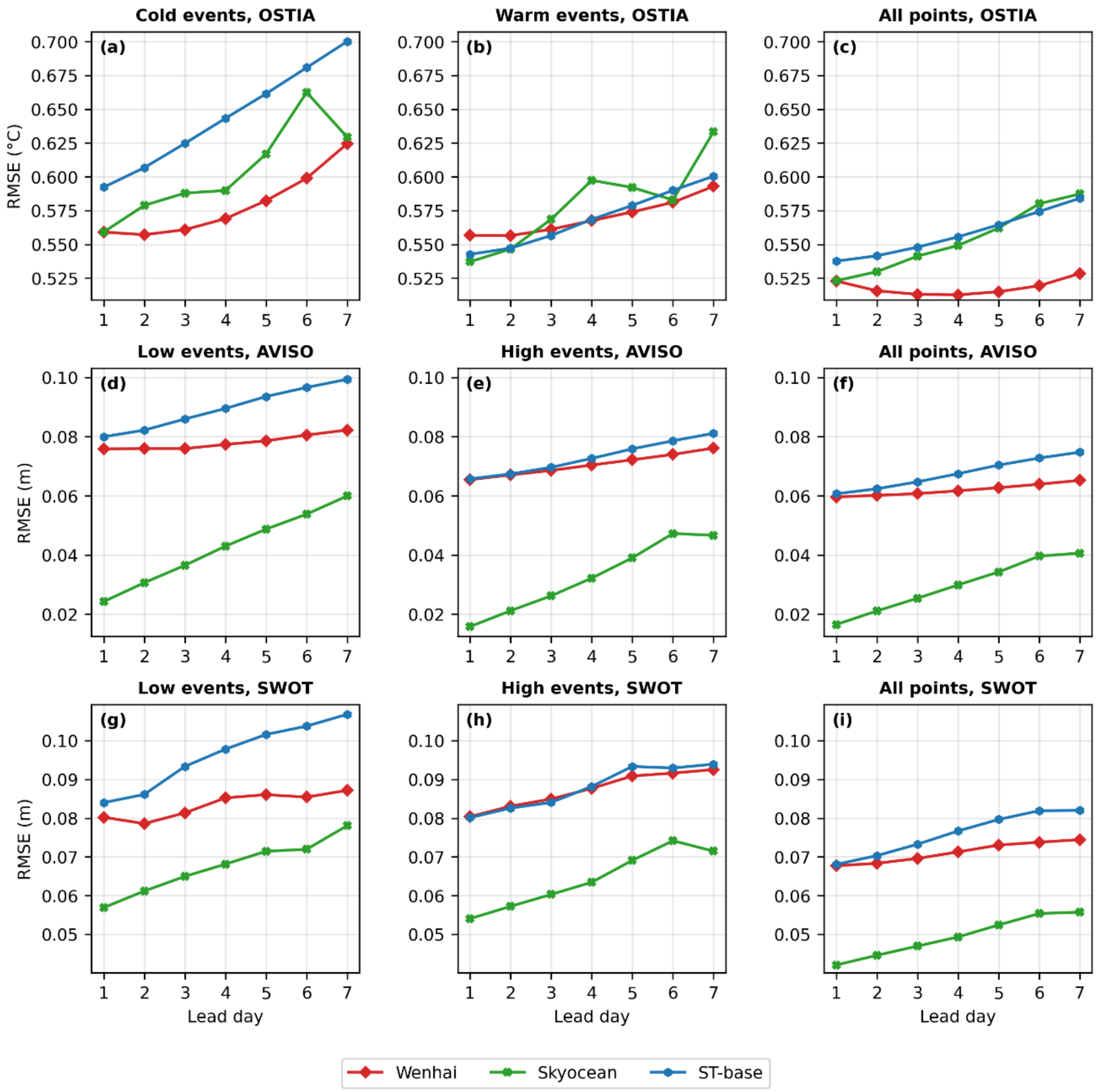

Figure 3. RMSE for extreme values and all valid samples with lead day. Panels (a–c), panels (d–f) and panels (g–i) show SST against OSTIA, SLA against AVISO L4, and SLA against SWOT L3, respectively. Lower- and upper-tail extremes are defined using the 10th and 90th percentile thresholds, respectively. These percentiles are derived from historical OSTIA SST and AVISO L4 SLA data.

Forecast errors are also asymmetric between the two tails. Negative extremes (cold SST and low SLA) generally have larger RMSEs than the positive extremes (warm SST and high SLA). The RMSEs of the three forecasts are also more similar for warm SST extremes than for cold SST extremes, showing that the differences in forecast errors depend on which tail is considered. Extreme-event RMSE therefore provides tail-specific information about forecast errors that cannot be revealed by full-sample RMSE alone.

SLA RMSEs also depend on the reference used for evaluation. For Skyocean, RMSE for extreme events increases substantially when the reference is changed from AVISO L4 to SWOT L3. AVISO L4 is used during Skyocean training and inference. Comparing the forecasts against AVISO L4 and SWOT L3 therefore helps determine whether the low RMSE is specific to AVISO or persists against an independent reference (Zu et al., 2025).

### 4.3.2 Power spectral density

PSD characterizes how variance is distributed across spatial scales. Figure 4 presents only the wavelength range shorter than 150 km. At lead day 1, the large-scale SST spectra of all three forecasts are generally close to the OSTIA spectrum (not shown), but their spectral energy is lower than that of OSTIA at wavelengths of approximately 30–70 km. By lead day 7, this energy deficit becomes more pronounced.

The SLA spectra depend strongly on the reference. Relative to AVISO L4, the Skyocean spectrum is close to the reference, consistent with its small AVISO-based pointwise errors (Figure 3f), whereas Wenhai and ST-base present higher spectral energy than AVISO. These differences become more pronounced by lead day 7. Relative to SWOT L3, however, all three forecasts show less spectral energy than the reference. Skyocean has a pronounced energy deficit at fine scales, whereas ST-base retains spectral amplitudes closer to SWOT. Therefore, agreement with AVISO does not prove that the fine-scale variability observed by SWOT has been retained.

These results should be interpreted in light of the different spatial resolutions and sampling characteristics of the two reference datasets. AVISO L4 is a gridded product constructed by merging multi-mission altimeter observations, whereas SWOT L3 provides higher-resolution observations along wide swaths (Ballarotta et al., 2019; Morrow et al., 2019). Therefore, differences in metric values partly reflect differences in the spatial characteristics of the reference datasets, and direct comparisons of forecast errors across the two references should be made with caution.

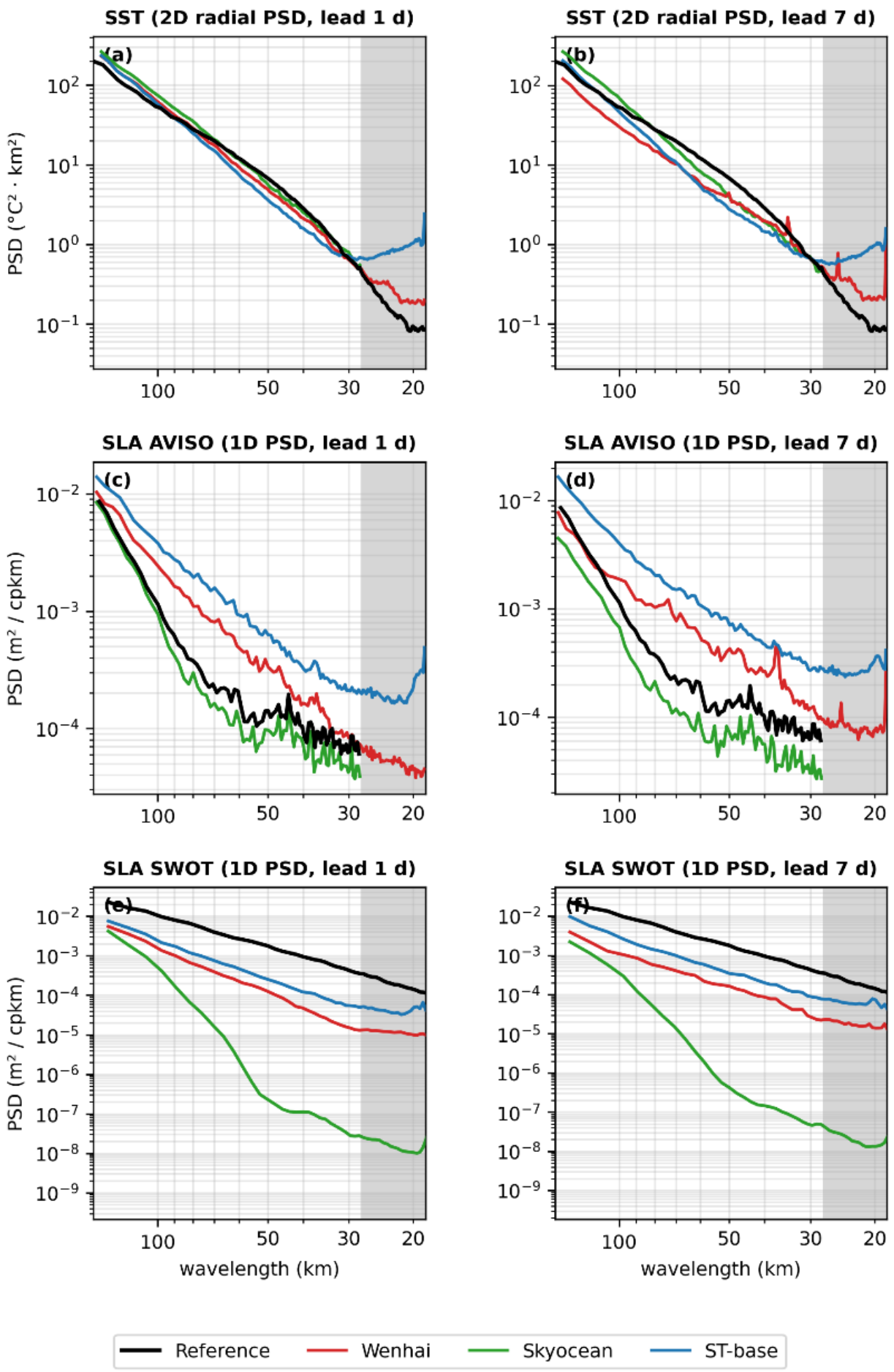

Figure 4. PSD of the forecasts and references at lead days 1 (left panel) and 7 (right panel). Gray shading denotes the short-wavelength range that is not resolved by all three forecast grids, and its left and right boundaries correspond to grid spacings of 1/8° and 1/12°, respectively.

### 4.3.3 Scale-decomposed forecast errors

Scale-decomposed RMSE reveals differences in forecast errors across spatial scales and reference datasets (Figure 5). For SST, Wenhai generally has the lowest RMSE at Scales A and B, whereas Skyocean performs best at Scale C (Figures 5a–c). For SLA against AVISO L4, Skyocean has the lowest RMSE across all three scales and lead times (Figures 5d–f). Against SWOT L3, Skyocean has the lowest RMSE at Scale A, Wenhai at Scale B, and the two show comparable RMSEs at Scale C (Figures 5g–i). These results cannot be fully captured by full-field RMSE alone.

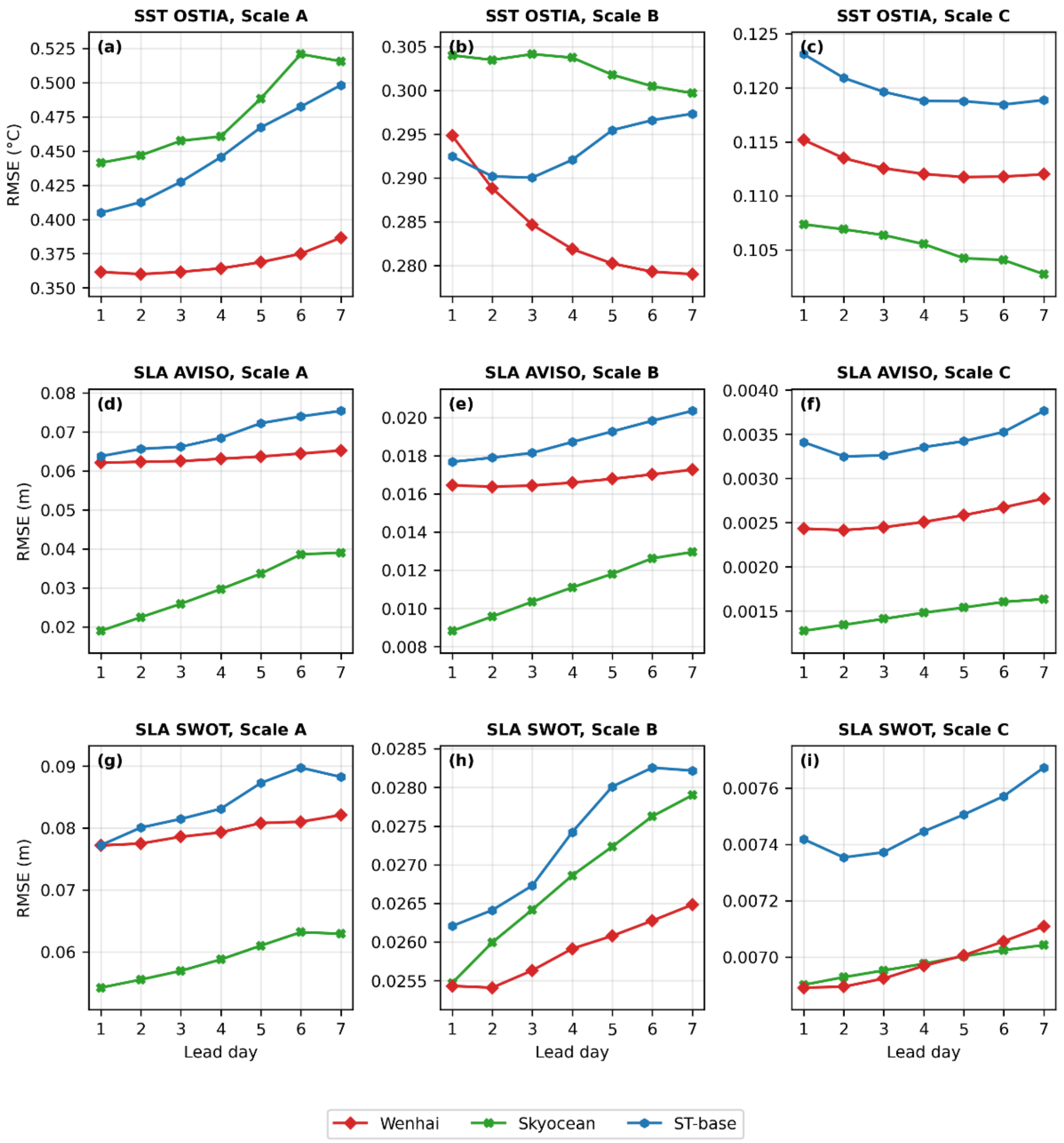

Figure 5. Scale-decomposed RMSE with lead time. Scales A (left panel), B (middle panel), and C (right panel) are defined as wavelength bands greater than 300 km, between 70 and 300 km, and less than 70 km, respectively.

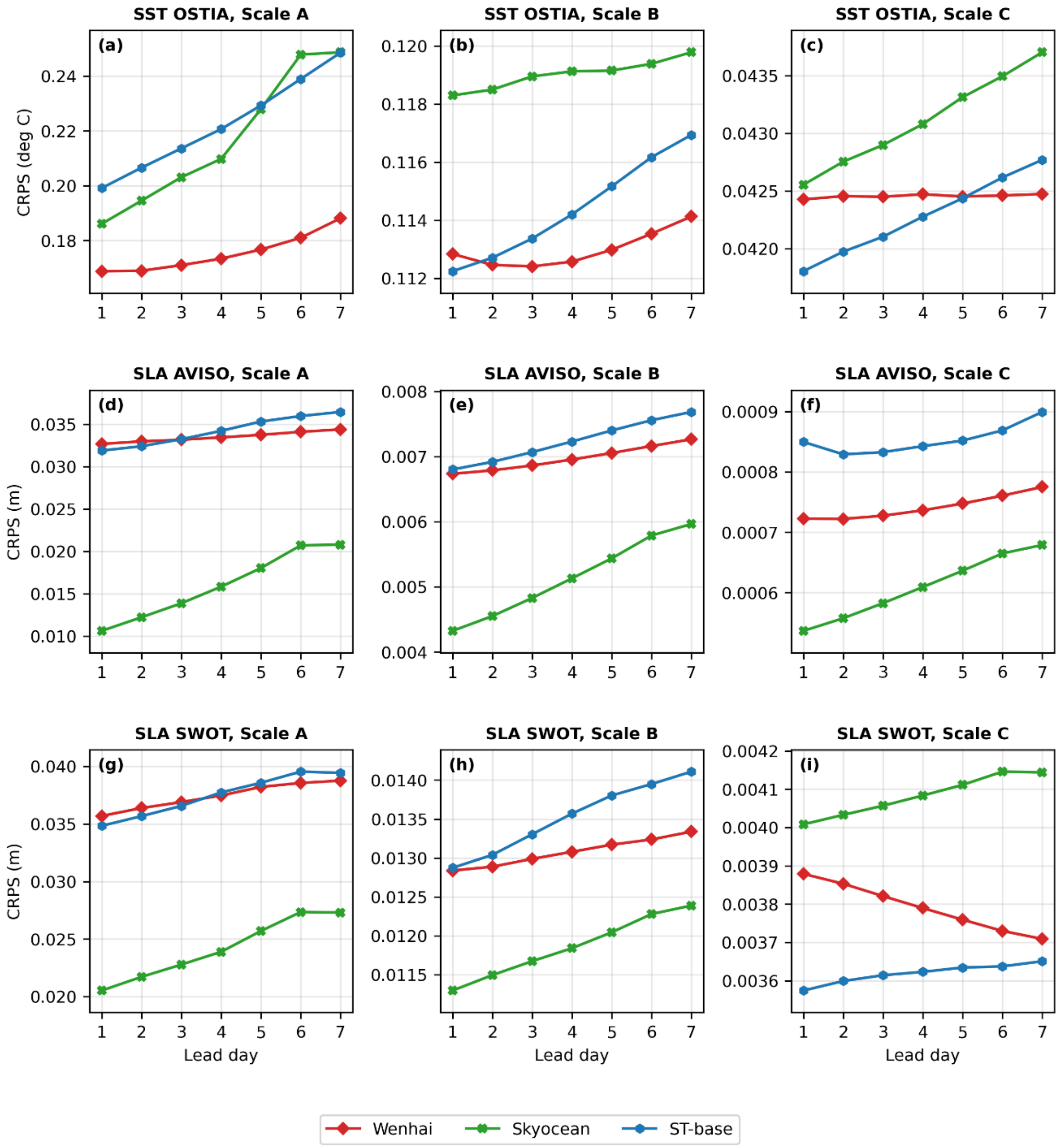

Figure 6. As in Figure 5, but for the scale-decomposed neighborhood CRPS.

Neighborhood CRPS complements RMSE because it is less sensitive to small spatial displacements of forecast structures with respect to the reference (Figure 6). For SST, the two metrics are broadly consistent at Scales A and B but differ at Scale C. The same fine-scale contrast appears for SLA against SWOT L3 (Figures 5i and 6i). At Scale C, ST-base has a large RMSE but a small CRPS and spectral amplitudes closer to SWOT (Figures 4e and 4f). This suggests that spatial displacement contributes to its fine-scale

RMSE. Skyocean has a smaller RMSE but a larger CRPS and a pronounced spectral deficit, suggesting spatial smoothing and reduced variance. Taken together, PSD, scale-decomposed RMSE and CRPS help distinguish displacement errors from errors associated with smoothed structures (Gilleland et al., 2009; Stein and Stoop, 2022).

For both SST and SLA, the magnitudes of scale-decomposed RMSE and neighborhood CRPS decrease from Scale A to Scale C. This pattern mainly reflects that ocean-state variance and forecast error concentrate at larger scales. Because large-scale errors dominate full-field RMSE, scale decomposition is necessary to assess errors in intermediate- and fine-scale structures separately. The scale dependence also has implications for model training: a model can achieve a low full-field loss by focusing on Scale A alone, which may explain why AI forecasts often exhibit smoothing of fine-scale structures.

## 5 Discussion

### 5.1 Methodological advances

Compared with existing benchmarks based on fixed historical datasets and overall pointwise errors, this study makes three methodological advances. First, multiple models are assessed through continuous 7-day rolling forecasts in a unified operational environment, allowing end-to-end performance to be evaluated under operational conditions rather than on fixed offline datasets. Second, the framework introduces physical-consistency metrics that characterize cross-variable balance relationships independently of reference datasets. Third, it combines extreme-event RMSE, PSD, scale-decomposed RMSE, and neighborhood CRPS to evaluate forecast performance for extreme values and at fine spatial scales, helping distinguish spatial smoothing from errors associated with the displacement of fine-scale structures. Together, these advances extend AI ocean forecast evaluation beyond numerical agreement with a reference and provide a basis for model intercomparison, model development, and operational evaluation.

### 5.2 Complementary evaluation dimensions for AI forecast trustworthiness

The three dimensions of the proposed evaluation framework provide complementary perspectives on forecast trustworthiness. Overall pointwise accuracy measures the overall agreement between forecasts and references across all valid samples. Physical consistency examines whether forecast variables remain mutually coherent with respect to dynamical balances, whereas distributional consistency evaluates forecast errors for the extreme values and fine-scale structures. This framework shifts evaluation from ranking based on a single metric toward assessment across three complementary dimensions, consistent with recent advances in AI forecast evaluation (Rasp et al., 2024; El Aouni et al., 2025b).

These three evaluation dimensions are complementary rather than interchangeable. The SLA results illustrate this complementarity: the forecast with the smallest pointwise RMSE exhibits a relatively large geostrophic imbalance and the greatest fine-scale variance deficit, whereas a larger pointwise error may correspond to the spatial displacement of fine-scale structures. Thus, favorable values of overall pointwise metrics do not necessarily imply physical or distributional consistency, and vice versa. Combining overall pointwise metrics with physical-consistency and distributional-consistency metrics helps distinguish spatial smoothing, structural displacement, and cross-variable inconsistency. Forecast trustworthiness should therefore be assessed by considering all three dimensions.

### 5.3 Implications for model development and operational implementation

Metrics from multiple evaluation dimensions can guide model development. Fine-scale variance deficiency may be reduced by supplementing pointwise losses with spectral or structural-similarity terms, whereas spatially tolerant objectives can reduce the double penalty associated with small displacements of eddies and fronts (Cui et al., 2025). Residuals derived from dynamical balances may likewise be formulated as physics-based regularization terms. These strategies are consistent with physics-informed and hybrid modeling approaches that incorporate scientific knowledge into deep learning frameworks (Raissi et al., 2019; Kashinath et al., 2021; Kochkov et al., 2024).

For operational implementation, models should be assessed before deployment using application-specific combinations of metrics rather than relying on a single metric. Overall pointwise accuracy is critical for applications concerned with the large-scale background state, whereas extreme-event services require tail-specific error metrics and applications involving eddies and fronts require scale-decomposed metrics. Physical-consistency metrics are particularly important for physical explanation. The framework can guide the design of scoring rules for future AI ocean forecasting competitions. Using metrics from all three dimensions would help avoid rewarding models that achieve lower RMSEs at the expense of physical coherence or fine-scale structure.

### 5.4 Limitations and future work

AI-based ocean forecasts mainly provide daily-mean forecasts and typically do not include vertical velocity. These limitations restrict the assessment of conservation properties, such as mass and potential vorticity conservation. Future work will require higher-frequency outputs and additional forecast variables to extend the framework to conservation metrics.

The present study also has several limitations. The geostrophic-balance metric is sensitive to the choice of deformation radius, filtering scale, and spatial weighting. PSD estimates near the model grid scale are affected by interpolation, sampling, and the effective resolution of the reference data. Further work is needed to assess the sensitivity of these metrics. In addition, this study covers only three forecasting systems, one region, a four-month period, and lead times of 1–7 days. Addressing these limitations will help make the framework more robust and broadly applicable to AI ocean forecasts.

## 6 Conclusions

This study develops a multidimensional evaluation framework for AI ocean forecasting, comprising overall pointwise accuracy, physical consistency, and distributional consistency. The main conclusions are as follows.

First, continuous rolling evaluation in a unified operational environment provides a more realistic assessment of forecast skill and generalization than offline evaluation on fixed historical datasets. By evaluating complete forecasting systems rather than neural network architectures alone, it better reflects their performance under operational conditions.

Second, favorable overall pointwise accuracy does not necessarily imply physical consistency. Small SLA errors can coexist with substantial geostrophic-balance residuals, while stratification stability may evolve differently from pointwise errors.

Third, favorable overall pointwise accuracy does not necessarily imply distributional consistency. Small overall pointwise errors can coexist with larger errors for extreme values and deficient fine-scale variance. Conversely, larger fine-scale errors may partly reflect the displacement of coherent structures rather than their poor representation.

## Open research

The Argo temperature and salinity profiles, surface drifter observations, and along-track altimetry data used for overall pointwise accuracy were obtained from the OceanPredict IV-TT class 4 website. The OSTIA SST, AVISO L4 SLA, and SWOT L3 observations used for distributional consistency are available from their respective data providers.

## Acknowledgments

We thank the OceanPredict Intercomparison and Validation Task Team (IV-TT) for providing the reference datasets used to evaluate overall pointwise accuracy. This work was supported by the National Natural Science Foundation of China (Grant Nos. 42375062 and 42676003).

## Conflict of Interest

The authors declare there are no conflicts of interest for this manuscript.